\documentclass[reprint,showpacs,amsmath,amssymb,aps,prb]{revtex4-1}
\usepackage{graphicx}
\usepackage{dcolumn}
\usepackage{bm}
\usepackage{hyperref}

\usepackage[usenames,dvipsnames,table]{xcolor}
\usepackage[version=3]{mhchem} 
\usepackage{braket}
\usepackage{multirow}
\usepackage{soul}

\definecolor{amaranth}{rgb}{0.9, 0.17, 0.31}
\definecolor{battleshipgrey}{rgb}{0.52, 0.52, 0.51}
\definecolor{pastelred}{rgb}{1.0, 0.41, 0.38}
\definecolor{mark}{rgb}{0.85, 0.9, 1}

\newcommand{\hlmath}[1]{\text{\hl{#1}}}

\hypersetup {
	hidelinks
}
\sethlcolor{mark}

\begin{document}

\title{Monte Carlo study of magnetic nanoparticles adsorbed \\on halloysite \ce{Al2Si2O5(OH)4} nanotubes}

\author{O. M. Sotnikov$^1$, V. V. Mazurenko$^1$, A. A. Katanin$^{2,1}$}
\affiliation{$^1$Theoretical Physics and Applied Mathematics Department, Ural Federal University, Mira Str. 19, 620002 Ekaterinburg, Russia\\
$^2$Institute of Metal Physics, Kovalevskaya str. 18, Ekaterinburg 620990, Russia}

\date{\today}

\begin{abstract}
We study properties of magnetic nanoparticles adsorbed on the halloysite surface. For that a distinct magnetic Hamiltonian with random distribution of spins on a cylindrical surface was solved by using a nonequilibrium Monte Carlo method. The parameters for our simulations: anisotropy constant, nanoparticle size distribution, saturated magnetization and geometrical parameters of the halloysite template were taken from recent experiments. We calculate the hysteresis loops and temperature dependence of the zero field cooling (ZFC) susceptibility, which maximum determines the blocking temperature. It is shown that the dipole-dipole interaction between nanoparticles moderately increases the blocking temperature
and weakly increases the coercive force.
The obtained hysteresis loops (e.g., the value of the  coercive force) for Ni nanoparticles are in reasonable agreement with the experimental data.
We also discuss the sensitivity of the hysteresis loops and ZFC susceptibilities to the change of anisotropy and dipole-dipole interaction, as well as the 3d-shell occupation of the metallic nanoparticles; in particular we predict larger coercive force for Fe, than for Ni nanoparticles.
\end{abstract}

\maketitle

\section{\label{sec:intro}Introduction}

Halloysite, having chemical composition \ce{Al2Si2O5(OH)4}, is a natural aluminosilicate clay compound that has multi-walled tubular morphology and, along with the natural sources, can be formed from kaolinite~\cite{kaol2halloy}. The  tube single wall consists of two layers: the outer \ce{SiO4} tetrahedral layer and the inner octahedral \ce{AlO6} layer. Typical length, inner and outer diameters of the individual halloysite nanotube are about 600-900 nm, 15 nm and 50 nm, respectively\cite{Lvov_review}. However, the diameter of the tubes can vary in the range 30-190 nm depending on the deposit~\cite{halloysite_review}. There are many applications of the halloysite were suggested, including sustained drug delivery~\cite{drug_delivery,drug_delivery_2}, bone cement improvement~\cite{bones}, anti-corrosion~\cite{anti-corrosion,anti-corrosion_2, anti-corrosion_3} and flame-retardant~\cite{flame-ret} agents carrying, nanoconfined catalysts~\cite{catalysts}, environment cleaning~\cite{spilling}, biological cells coating~\cite{cell-coat}, and others~\cite{Lvov_review}. The material is biocompatible and available at low price, which makes it very attractable for both industry and research.

Being a nonmagnetic material, halloysite nanotubes can be used as a cylindrical template that provides a high stability of the transition-metal nanoparticles at high temperatures. For example, the authors of Ref.~\onlinecite{Ni_halloy} reported on the fabrication of the cermet composite with nickel nanoparticles adsorbed on the halloysite via electroless plating. Importantly, the connection with halloysite surface prevents the oxidation of the nanoparticles during several months. It was also found that the inherent coercive force (iHc) of the cermet composite is much higher than those measured for bulk Ni, which is a clear indication of a strong magnetocrystalline anisotropy of the individual Ni-nanoparticles. The structural analysis revealed that the distribution of the nanoparticles is uniform and the mean diameter is about 25-30 nm. Taking into account the typical size of halloysite nanotubes, it yields about 150 single-domain nanoparticles in total adsorbed on each nanotube. Similar magnetic measurements were performed for cobalt nanoparticles on the halloysite \cite{Co_halloy}.
\begin{figure} [!h]
	\includegraphics[width=\columnwidth]{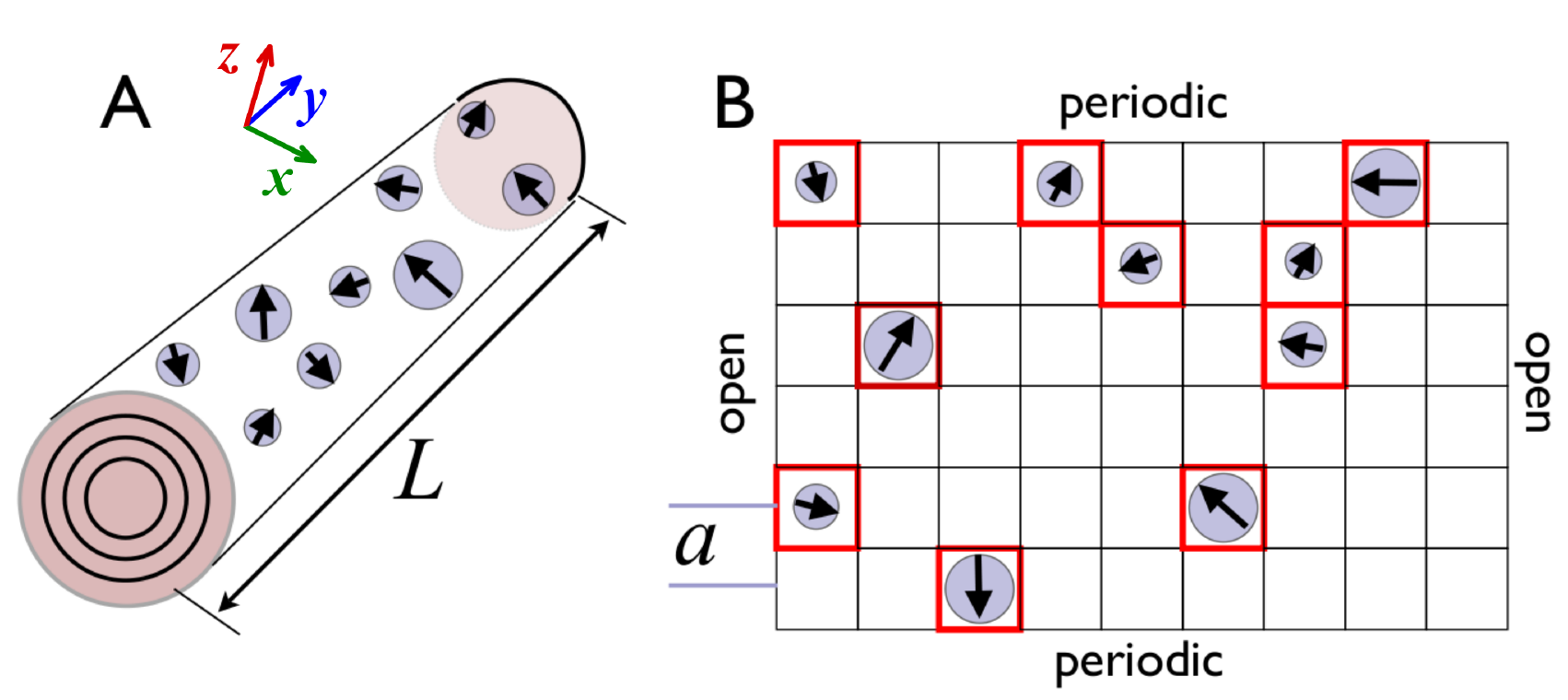}
	\caption{(Color online) (A) Schematic representation of halloysite nanotube of the length $L$ with adsorbed Ni nanoparticles of different sizes. (B) The corresponding unrolled lattice topology with the lattice constant $a$. Arrows inside circles denote the magnetic moments of the nanoparticles. It is important to note, that the cylindrical lattice has extra dimension along {\it z}-axis.}\label{fig:geom}
\end{figure}

Therefore, halloysite provides a natural substrate for studying physical properties of the magnetic nanoparticles, including the effects of their interaction. In this regard, an important question arises whether the known effects of the dipole-dipole interaction, such as superspin glass-like freezing, observed previously in system of interacting nanoparticles (see, e.g., Refs. \onlinecite{Freez1,Freez2,Freez21,Freez22,Freez3,Freez4,Freez5}), persist on the cylindrical geometry.  

The theoretical description of the cermet composite can be performed on different levels. For instance, to describe the formation of the individual Ni nanoparticles of different sizes as well as their anisotropies one needs to take into account the electronic structure of nickel clusters and magnetic interactions between nickel atoms.  These problems can be solved by means of the density functional theory methods and spin Hamiltonian approaches.\cite{Mn12}  In turn, the change of the physical properties of the Ni nanoparticles due to the contact with the halloysite surface is another complicated computational problem that requires the implementation of the Anderson-model-based methods.\cite{CuN, FePt}  

Here we consider higher level modeling, which treats magnetic nanoparticles adsorbed on the cylindrical surface and interacting with each other via dipole-dipole interaction. We focus on the theoretical description of the effects of disorder and the dipole-dipole interaction on the magnetization curves, comparing the results to the  experimental ones for Ni nanoparticles on the halloysite template. Apart from treatment of random positions and randomly-oriented anisotropy of nanoparticles, addressed in previous studies\cite{kech2002,kech2008}, 
we also consider some distribution of anisotropy values, which occurs due to different sizes of nanoparticles, and model positions of nanoparticles on the cylindrical surface.  We also assume randomly oriented (equal for all nanoparticles) magnetic field while averaging over disorder, as it occurs because of random orientation of nanotubes. The model was solved by means of the Monte Carlo approach.
It is essential, that the cylindrical lattice morphology (see Fig. 1) yields 
additional components of the dipole-dipole interaction, which are absent in flat case because of nanoparticles entirely belonging to a single plane (some components of the exchange interaction are also changed). Such a geometry with open boundary conditions therefore differs from the pure two-dimensional analogs,\cite{kech2002,kech2008} since it yields an additional source of the anisotropy for the 
nanoparticles.

Our main findings are the following. The coercive force and especially the blocking temperature are enhanced by the dipole-dipole interaction between nanoparticles on the cylindical surface. For nickel nanoparticles we obtain hysteresis loop in reasonable agreement with the experimental data. For iron nanoparticles adsorbed on the halloysite we predict a higher coercive force than that for nickel nanoparticles, while cobalt nanoparticles at room temperature are expected to be superparamagnetic. 

\section{\label{sec:methods}Method}
{\it Spin Hamiltonian.} To describe the cermet composite of Ni nanoparticles on the halloysite surface we introduce the following spin Hamiltonian:
\begin{equation}
 \label{eq:ham}
\mathcal{H} = \sum_{i<j} \sum_{\alpha\beta}J_{ij}^{\alpha \beta}S_i^{\alpha}S_j^{\beta} - \sum_ik_i\,(\mathbf{n}_i\cdot\mathbf{S}_i)^2
     - \sum_i\mathbf{h}_i\cdot\mathbf{S}_i,
\end{equation}
where $\alpha$($\beta$) = $x,y,z$ and $\mathbf{S}_{i}$ is a unit vector along the direction of the spin of the {\it i}-th nanoparticle, which are randomly distributed on the cylindrical surface.
$k_i = K_1V_i = {\pi}K_1d_i^3/{6}$ is the single-site anisotropy  ($K_1$ is the anisotropy constant) and $d_{i}$ is the diameter of the {\it i}-th nanoparticle, which, in according with the experiment for nickel nanoparticles, \cite{Ni_halloy} has a normal distribution with the mean diameter $\langle d \rangle$ = 25-30 nm. $\mathbf{h}_i = \mu_i\mathbf{H}$ denotes the energy contribution from external magnetic field ${\mathbf H}$, where $\mu_i = M_sV_i$ is the magnetic moment of the {\it i}-th particle, $M_s$ represents the saturation magnetization of the particles per unit volume; the direction of the magnetic field is randomly chosen during disorder averaging. 

Previous theoretical studies of the magnetic properties of the dipolar interacting nanoparticles assemblies (for instance, see Refs. \onlinecite{kech2002,kech2008,Ni_np_12})   concentrated on the effect of the random orientation of the uniaxial anisotropy, keeping its value fixed, which corresponds to the case of the nanoparticles of the identical size. However, experimentally there is always some distribution of sizes of nanoparticles (see, e.g. Ref. \onlinecite{Ni_halloy} for the nickel nanoparticles on halloysite). Therefore, in our case
the anisotropy term is characterized not only by the random orientation of the easy axis but also by a random value of the anisotropy energy related to the size of the particle. 

In turn, the dipole-dipole interaction tensor is given by 
\begin{equation}
 \label{eq:j}
 J_{ij}^{\alpha \beta} = J_0\frac{1}{R_{ij}^3}(\delta_{\alpha \beta}-\frac{3R_{ij}^{\alpha}R_{ij}^{\beta}}{R_{ij}^2}),
\end{equation}
where $J_0= \langle\mu\rangle^2 = M_s^2\langle{}V\rangle^2$, 
$R_{ij}^{\alpha}$ is the component of the vector pointing from \textit{i}-th to \textit{j}-th site, and $\delta_{\alpha \beta}$ is the Kroneker delta symbol. The calculation of the dipole-dipole interaction tensor is simpler in the cartesian coordinate system, which was used in our modeling. The curvature of the tubes yields changes in the components of the dipole exchange interaction tensor $J_{ij}^{\alpha \beta}$ in comparison to the flat case (in particular, some components, which are absent in the flat case appear), as discussed in Appendix A.

{\it Definition of the model parameters.} In our approach we should specify the values of the parameters for the model Hamiltonian, namely anisotropy constant $K_1$ and saturated magnetization $M_s$. They can be estimated from magnetic measurements; Table \ref{tab:parameters} summarizes several values known from literature for nanoparticles of different type and size. While the average sizes of the iron and nickel nanoparticles are close to each other, the anisotropy constant for Ni nanoparticles is about 20 times larger than that in the case of iron, possibly due to the difference in the spin-orbit coupling, which is stronger in systems with nickel than that with iron. \cite{wafm} In turn, the larger values of the saturated magnetization of iron nanoparticles can be explained by the proximity of the Fe $3d$ shell to the half-filling. Based on these experimental results we define the model parameters for our simulations. They are presented in Table II. 

\begin{figure}
	\includegraphics[width=0.9\columnwidth]{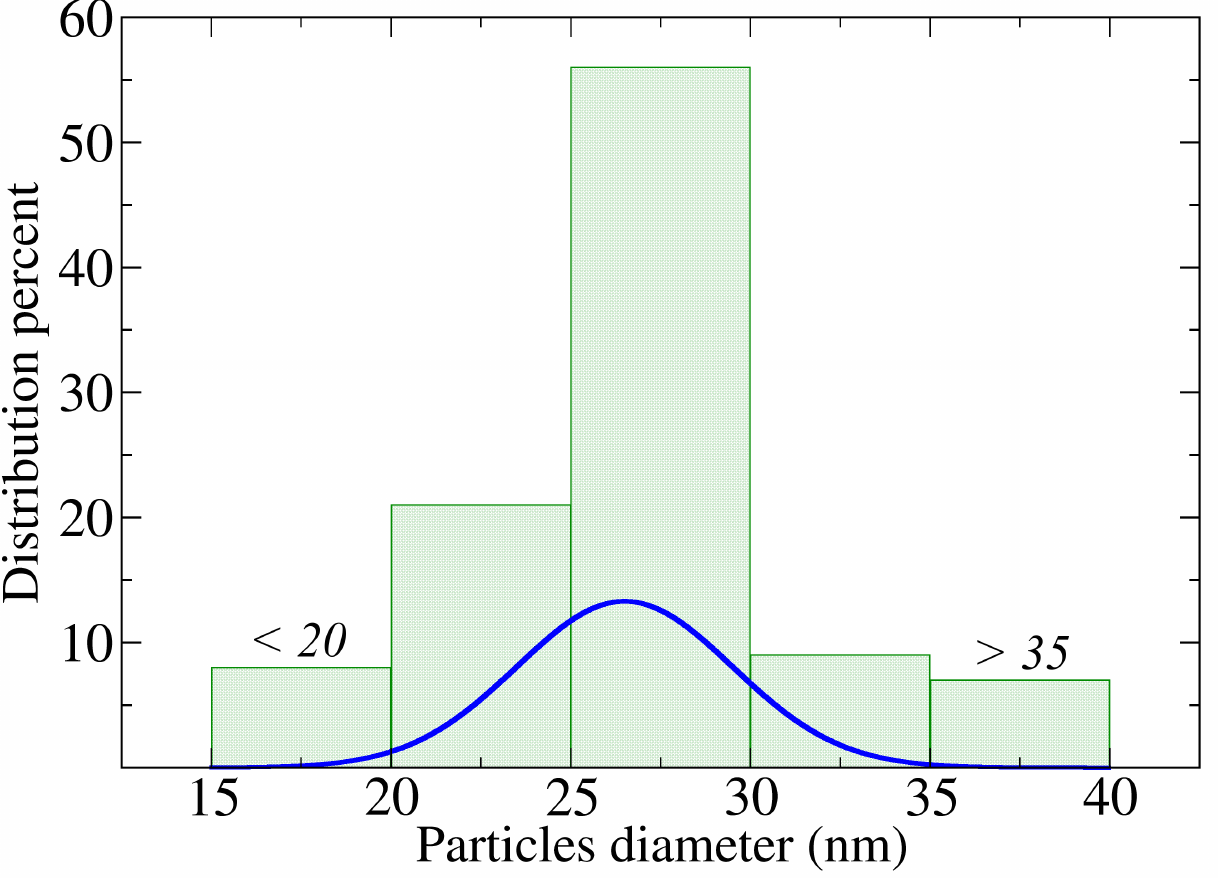}
	\caption{(Color online) Experimental size distribution of the nickel nanoparticles deposited on the halloysite. The data are taken from Ref.\onlinecite{Ni_halloy}. The solid line represents theoretical fitting used in this work.}\label{fig:dist}
\end{figure}

To define the geometry of our model, we construct a square lattice, rolled into cylinder, corresponding to a halloysite nanotube, observed in experiments~\cite{Ni_halloy}, with the diameter of 170 nm and length of 1000 nm. The lattice is introduced for calculation purposes only, to avoid an overlap of the neighbouring nanoparticles. For Ni nanoparticles the corresponding lattice parameter is chosen to be equal to 32 nm, somewhat larger the average diameter of adsorbed nanoparticles 26.5 nm. Sizes of particles were distributed according to the normal law with the standard deviation equal to 3 nm (see Fig.~\ref{fig:dist}). The concentration of nanoparticles has been set to 0.5, which is close to that observed in experiments,~\cite{Ni_halloy} and corresponds to 264 occupied sites in total. Each occupied site is associated with a randomly oriented 
anisotropy axis $\mathbf{n}_i$ and normally distributed size $d_i$ of the particles. Similar consideration was used for Fe and Co nanoparticles.
We note that because of smaller size, calculations for Co nanoparticles involve many more particles than simulations for Fe and Ni. For instance, for the same geometry described above the system contains 2860 cobalt nanoparticles. To reduce the computational time the tube diameter for the system with Co was chosen to be 90 nm.

\begin{table}
\caption[Bset]{\label{tab:parameters} List of the magnetic quantities determined from the experiments on iron, nickel and cobalt nanoparticles. $M_s$, $K_1$ and iHc are the saturated magnetization, anisotropy constant and coercive force, respectively. The row denoted with asterisks describes Ni nanoparticles adsorbed on the halloysite outer surface and treated at 573 K.}
\renewcommand{\arraystretch}{1.7}
\begin{tabular}{ccccccc}
\hline\hline
Ion & $T$ (K) & $\langle{}d\rangle$ (nm) & $ K_1 $ (erg/cc) & $ M_s $ (emu/g) & iHc (Oe) & [Ref.] \\
\hline
Ni & 300 & 23.5 & $2.4\cdot{}10^5$ & 36.4 & 190 & [\onlinecite{Ni_np_10}] \\
Ni$^*$ & 298 & 26.5 & \--- & 31.0 & 112.8 & [\onlinecite{Ni_halloy}] \\\hline
Co & 3 & 5.8 & $2.5\cdot{}10^6$ & 110 & 700 & [\onlinecite{russier2000}] \\
Co & 300 & 3-7 & \--- & 60.25 & 580.72 & [\onlinecite{Co_halloy}] \\\hline
Fe & 298 & 25 & $2.1\cdot{}10^6$ & 216.1 & 250 & [\onlinecite{Fe_np}] \\
\hline\hline
\end{tabular}
\end{table}

\begin{table}
\caption[Bset]{\label{tab:parameters_sim} Simulation parameters for different types of nanoparticles used in this work. $\braket{d}$, $\sigma$ and $a$ are the size distribution mean, standard deviation and lattice constant of the square cylindrical surface, respectively. }
\renewcommand{\arraystretch}{1.7}
\begin{tabular}{cccccc}
\hline
 Ion & $\braket{d}$\,(nm) & $\sigma$\,(nm) & $a$\,(nm) & $M_s$\,(emu/cc) & $K_1$\,(erg/cc) \\
 \hline
 Fe & 25 & 3 & 30 &  1707 & $2\cdot{}10^6$ \\
 Co & 5.8 & 0.3 & 7 & 1000 & $2.5\cdot{}10^6$ \\
 Ni & 26.5 & 3 & 32 & 276.2 & $1\cdot{}10^5$ \\
\hline
\end{tabular}
\end{table}

\begin{figure}[h]
	\includegraphics[width=\columnwidth]{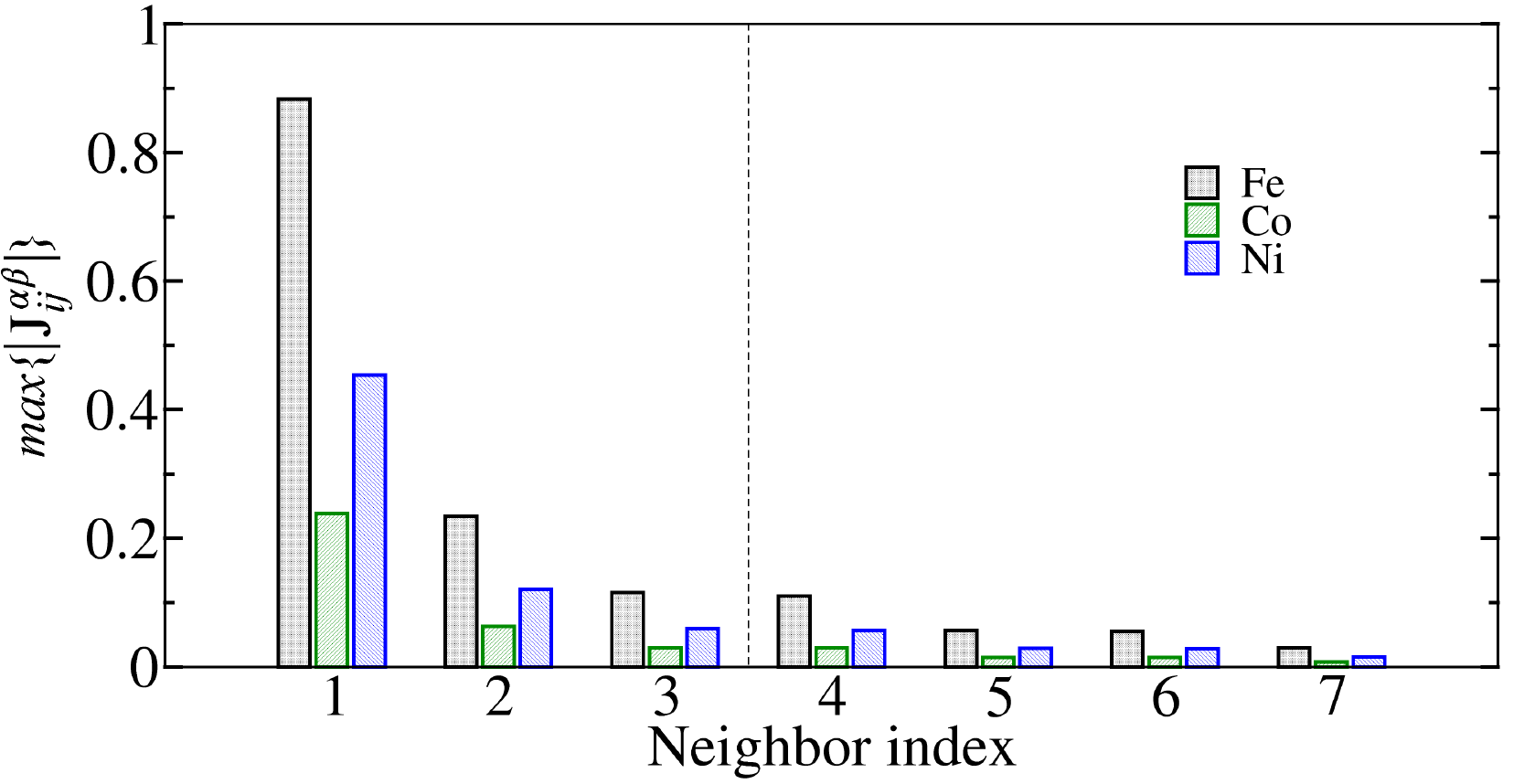}
	\caption{(Color online) Maximal values of the dipole-dipole interaction tensor $J_{ij}^{\alpha\beta}$ components depending on the neighbor index for Fe, Co and Ni nanoparticles. The values of the tensor are normalized by the mean value of the anisotropy energy $ \langle k \rangle ={\pi}K_1\langle d \rangle ^3/{6}$.}\label{fig:exch}
\end{figure}

By using the parameters of nanoparticles from 
Table \ref{tab:parameters_sim} we have calculated the typical dependence of the dipole-dipole interaction on the distance between different particles on the halloysite surface. The results are presented in Fig. \ref{fig:exch}. One can see that the elements of the dipole-dipole interaction tensor $J^{\alpha \beta}_{ij}$ strongly decay with distance. The interactions up to the third neighbors give the strongest contributions to the total energy of the system. Thus, we chose the largest coordination number $n_c = 3$ to reduce the computational time; we have verified that farther neighbors do not influence results appreciably.

{\it Solver}. To simulate the magnetization curves we solve the constructed spin model by means of the Metropolis algorithm with solid angle restriction (SAR) scheme \cite{bahiana2004} for classic Monte Carlo simulations. For initial system state preparation a heat bath~\cite{Landau} algorithm was used.
 
 One of the aims of the present paper is to reproduce the experimentally measured hysteresis loops, which requires a careful determination of the corresponding Monte Carlo parameters in accordance with conditions of the real experiment. 
 The correct  choice of the Monte Carlo simulation parameters (temperature- and field steps $\Delta{}T$ and $\Delta H$, solid angle restriction value $\Delta\Omega$, and the number of sweeps $N_{\rm SAR}$ per temperature or field change) can be based on the fitting of the magnetization curves.
Basing on the results of calculations described in Appendix~\ref{sec:appendix}, we determined, that $ \Delta{}H = 10$ Oe, $ \Delta{}T = 70$ K and $ N_{SAR} =$ 2000 sweeps are appropriate for angle restriction $ \Delta{}\Omega =$ 0.7 radian. 

\section{\label{sec:simulation}Simulation results}

\begin{figure}[t]
 	\includegraphics[width=\columnwidth]{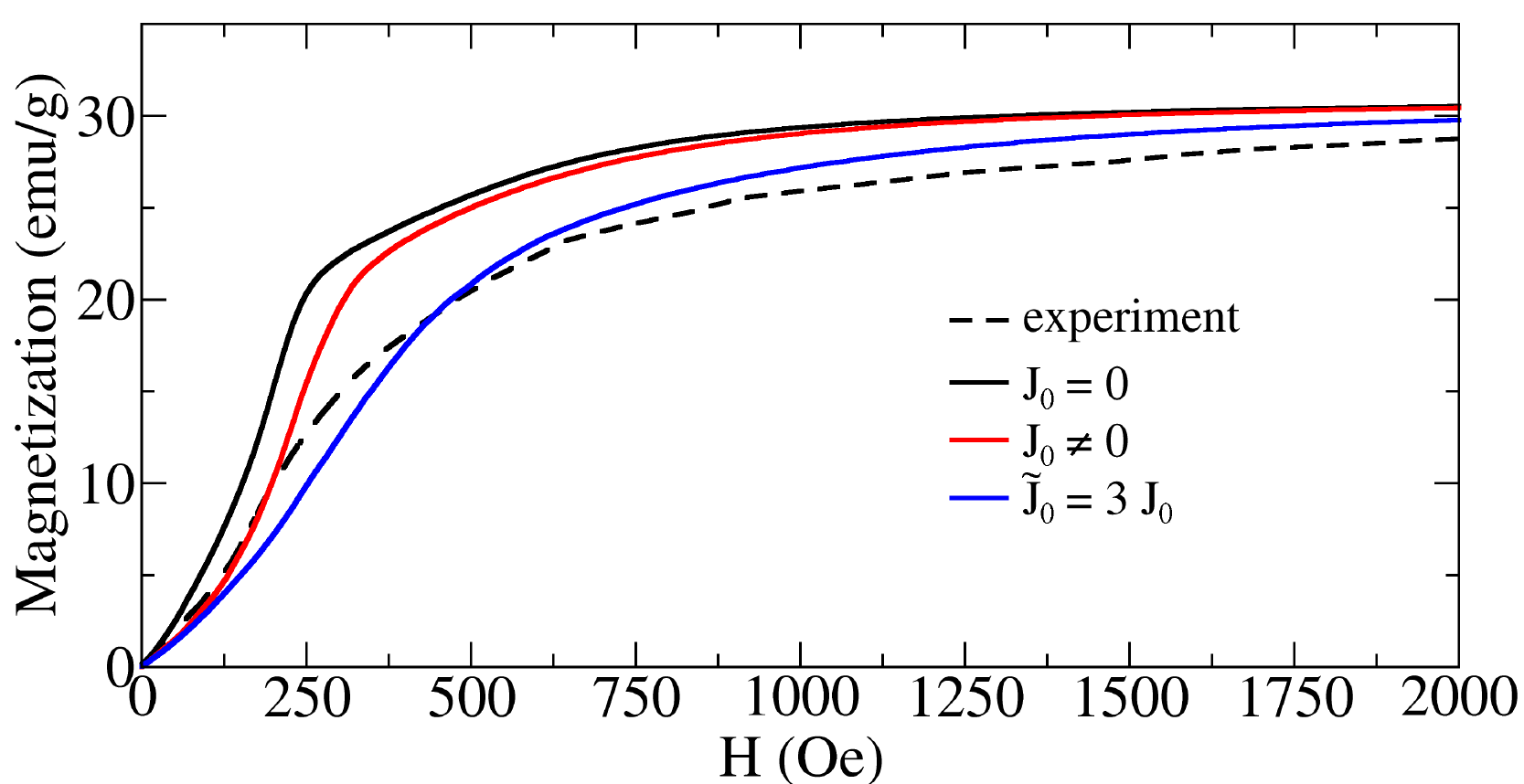}
 	\caption{(Color online) Comparison  of the experimental (dashed line) and calculated magnetization curves obtained at 300 K for the nickel cermet composite. The experimental data were taken from Ref.\onlinecite{Ni_halloy} 
 	}\label{fig:m573K}
 \end{figure}

\textit{Magnetization curve}.  In this section we discuss the results of simulation procedures. We consider first  magnetization curve; Fig.~\ref{fig:m573K} represents the results of calculation with chosen non-equilibrium Monte Carlo parameters. Although the small-field behavior is correctly reproduced, one can note, that the resulting slope of the magnetization curve at higher fields is still larger than the one experimentally observed.
The difference with the experimental data may originate from the underestimate of the interparticle interaction, which results from using the underlying square lattice, such that the interparticle distance is limited by the lattice constant. Indeed, using the larger value $ \tilde{J}_0 = 3J_0$, which corresponds to the $\sqrt[3]{3}$ times smaller average interparticle distance, we obtain better agreement with the experimental data.

\textit{ZFC magnetization curves.}
To study in more details the effect of the dipole-dipole interaction on the blocking temperature and a possibility of spin glass formation, we consider ZFC magnetization. 
The blocking temperature, which is the temperature of the transition to a superparamagnetic state~\cite{Np_Review}, can be determined from the peak of ZFC magnetization curves. 
The pronounced increase of this temperature with increasing concentration of the nanoparticles was previously observed in systems of interacting nanoparticles and interpreted as a signature of the superspin glass freezing, see, e.g., Refs. \onlinecite{Freez1,Freez2,Freez21,Freez3,Freez5}. 

\begin{figure}
	\includegraphics[width=\columnwidth]{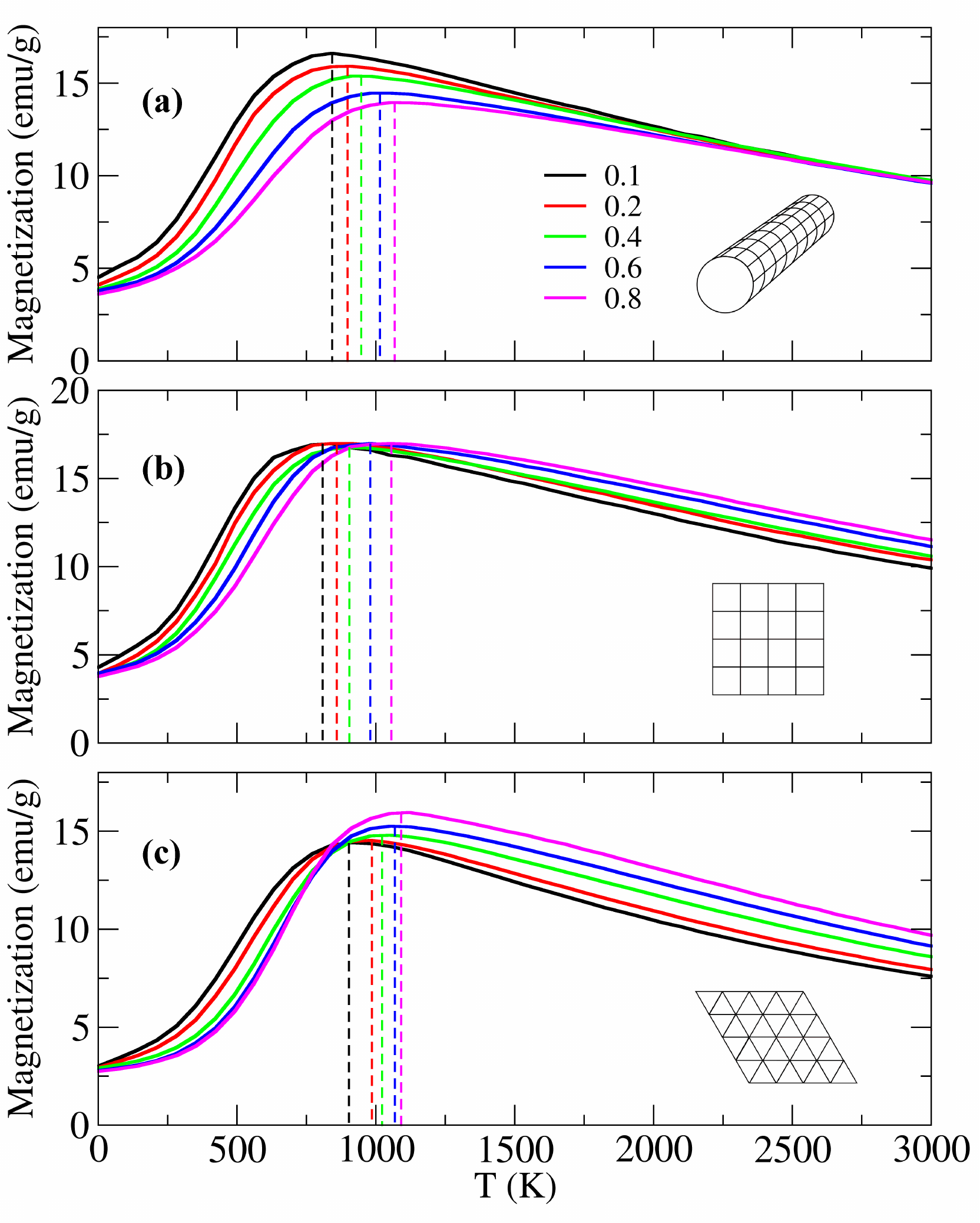}
	\caption{(Color online) 
		Calculated ZFC magnetization curves for Ni nanoparticles on (a) cylindric square, (b) flat square and (c) flat triangular lattices at different degree of the surface coverage. The in-surface field for square and triangular plane lattices was equal to 140 Oe and 100 Oe, respectively, and aligned along the lattice translation vector. Higher coating concentrations yield larger blocking temperatures.}\label{fig:zfc_Ni}
\end{figure}

\begin{figure}
	\includegraphics[width=\columnwidth]{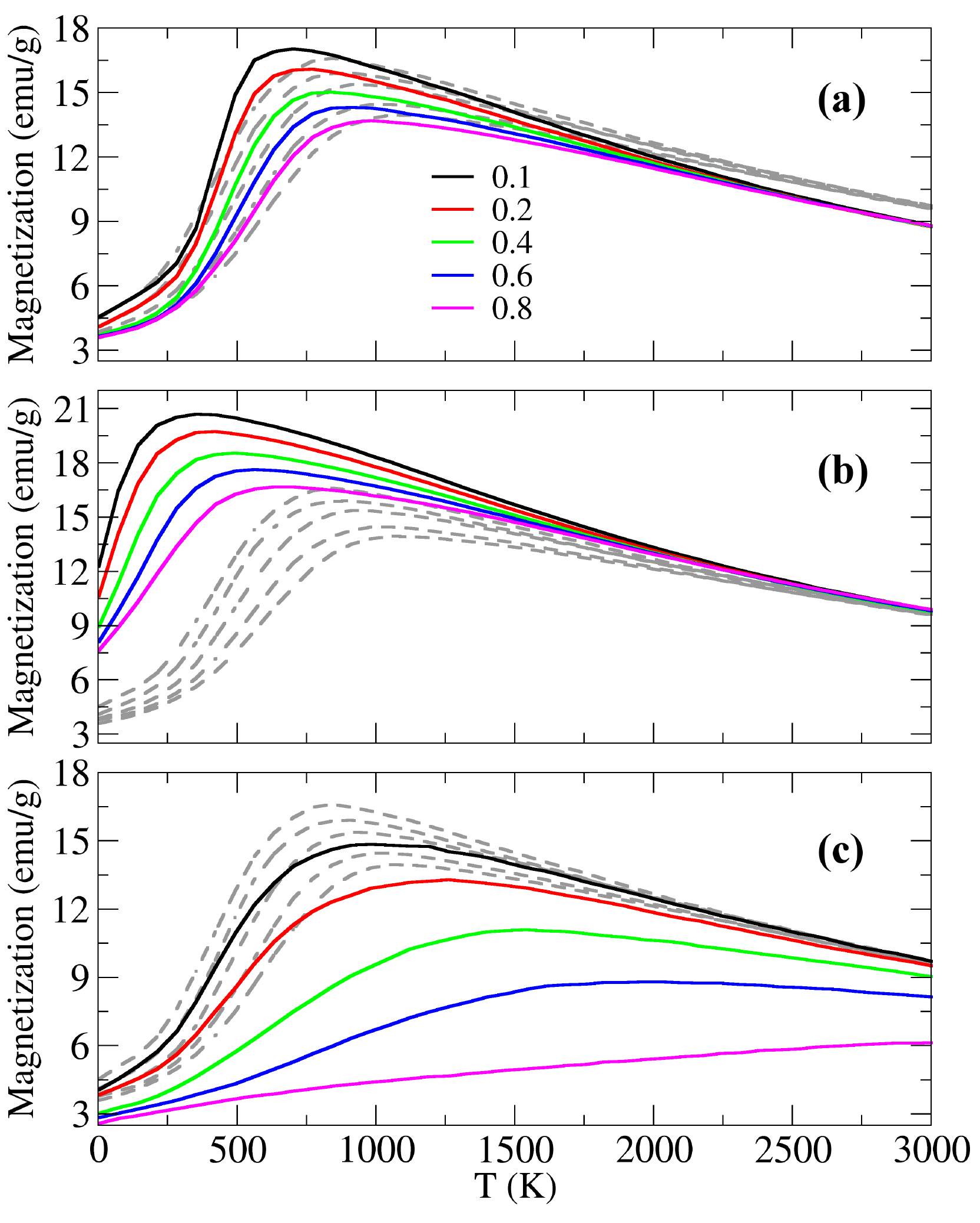}
	\caption{(Color online) 
		The influence of model parameters on the resulting ZFC curves. Dashed gray lines represent data with parameters for Ni. Colored lines denote (a) case of fixed particles size $ k_i = \braket{k} $, (b) case of decreased anisotropy constant $ \tilde{K}_1 = 0.5 K_1 $, (c) case of increased dipole-dipole interaction constant $ \tilde{J}_0 = 3J_0 $.}\label{fig:zfc_impact}
\end{figure}

To study the influence of the nanoparticle concentration on the magnetic properties, we calculated a number of the ZFC magnetization curves with different coating concentrations. Figure~\ref{fig:zfc_Ni}(a) shows the resulting ZFC magnetization curves for Ni nanoparticles. 
It is clear, that in the case of the non-interacting particles the ZFC magnetization curves do not depend on the number of the particles involved in the modeling. Including 
dipole-dipole interaction term in the Hamiltonian~\eqref{eq:ham}, we observe the pronounced increase of the blocking temperature $T_b$ with increasing of coating concentration, which is a clear indication of the interparticle interaction effect. This behavior is similar to previous results of simulation of the nanoparticles on the flat square lattice ~\cite{kech2002}, where it was shown that $T_b$ increases with decrease of the interparticle distance. 

Differently from the previous results for the flat square lattice\cite{kech2002}, the height of the ZFC susceptibility peak slightly decreases with increasing coating concentration and constitutes about $0.53 M_s$ for small surface coverage and $0.45 M_s$ for high concentration of the particles. Since the dependence of ZFC curves on the concentration is entirely due to the dipole-dipole interaction term of the Hamiltonian, the decrease of the peak is attributed to a specific geometry, which has a significant impact on the coupling term. It is important, that Ni-coated nanotubes in a macroscopic sample have random directions. In our theoretical approach we reproduce it by choosing a random direction of the magnetic field for each distribution of the nanoparticles. Therefore, it is impossible to choose a fixed direction of the external field relatively to a single tube geometry as in the case of flat lattices. 
In order to demonstrate the impact of such a treatment, we calculated ZFC magnetization curves for Ni nanoparticles on flat square lattice with the fixed in-plane direction of the magnetic field. The result of this simulation is represented in Fig.~\ref{fig:zfc_Ni}~(b). It is clearly seen, that in this case the peak is not sensitive to the coating concentration and constitutes about $0.54 M_s$.
We have also carried out the simulation for Ni nanoparticles distributed on the flat triangular lattice with magnetization field $H = 100$ Oe (Fig.~\ref{fig:zfc_Ni}~(c)). One can see that in this case the peak height slightly increases with growing concentration of the nanoparticles. In all considered cases, we see sizable effect of the dipole-dipole interaction, which results in the pronounced shift of the maxima of ZFC curves with the concentration of nanoparticles.

In Fig.~\ref{fig:zfc_impact} we consider the effect of nanoparticles size distribution, as well as changing the anisotropy and dipole-dipole interaction parameters. 
Considering sizes of nanoparticles 
equal (see Fig.~\ref{fig:zfc_impact}a) yields comparable shift of blocking temperatures to the results of Fig.~\ref{fig:zfc_Ni}, but stronger suppression of the maximum of ZFC curve at larger nanoparticle concentration.
It is important to note, that the impact of nanoparticles size distribution can be 
more pronounced 
without using the auxiliary lattice, since in this case different sizes of particles also lead to change of interparticle distances affecting the contributions from dipole-dipole interaction term. Lowering the anisotropy constant $K_1$, one observes corresponding decrease of the blocking temperature (which is expected to be proportional to the anisotropy for the system of non-interacting nanoparticles, see, e.g., Ref. \onlinecite{Np_Review}), as shown in Fig.~\ref{fig:zfc_impact}(b). However, variation of anisotropy does not change qualitatively the dependence of blocking temperature on concentration of nanoparticles. On the other hand, with increase of the dipole-dipole interaction (Fig.~\ref{fig:zfc_impact} c) we observe the shift of the maximum with the particle concentration, which is even more pronounced comparing to $\tilde{J}_0 = J_0 $ case, which clearly shows a possibility of realization of spin-glass state.

\begin{figure}
	\includegraphics[width=\columnwidth]{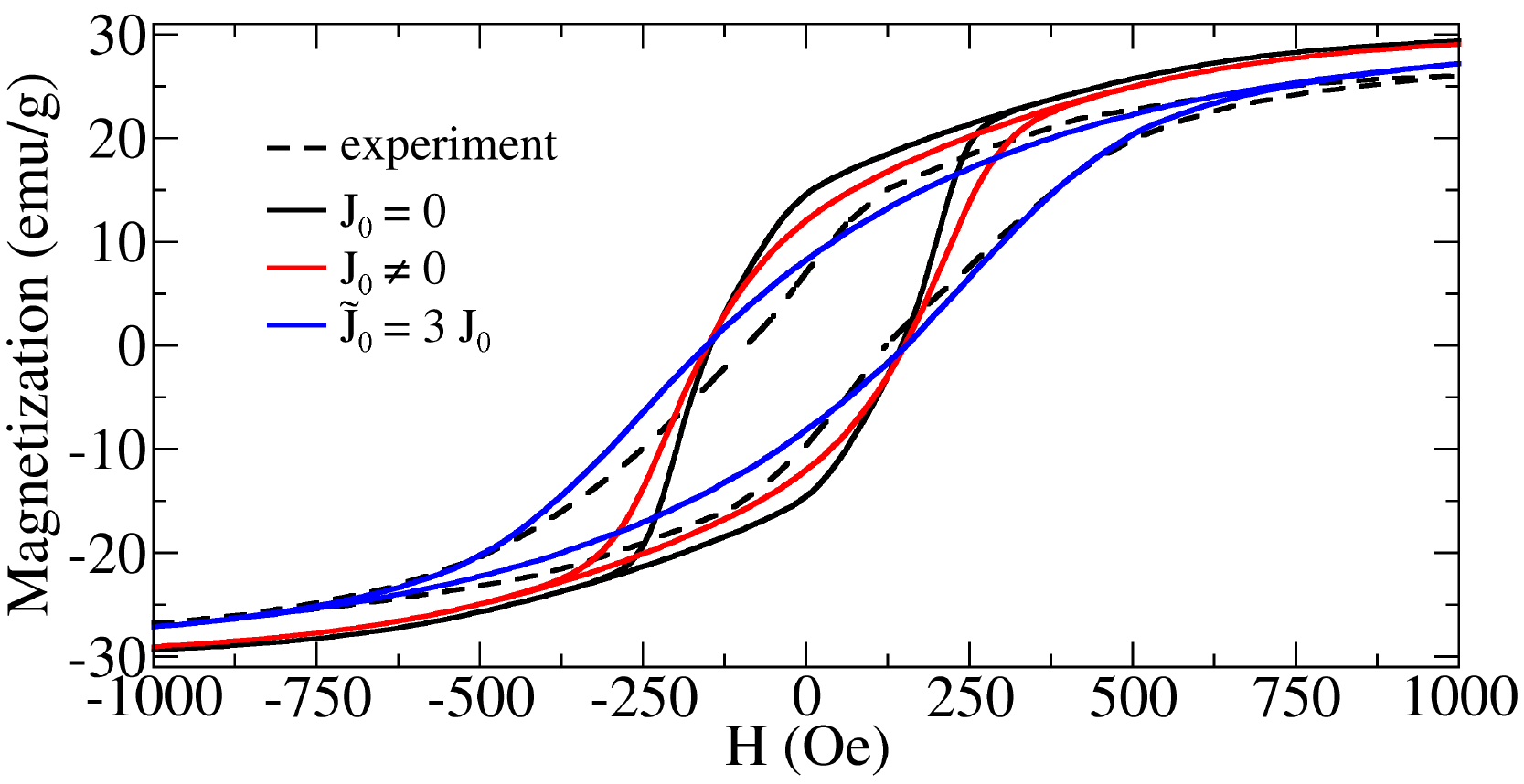}
	\caption{
    (Color online) Comparison of the experimental (dashed line) and calculated hysteresis loops obtained at 300 K for the nickel cermet composite. The experimental data were taken from Ref.\onlinecite{Ni_halloy}.}\label{fig:h573K}
\end{figure}

\textit{Hysteresis loops.} The corresponding results of calculation of hysteresis loop with and without dipole interaction for the parameters, corresponding to Ni nanoparticles, are presented in Fig.~\ref{fig:h573K}. As well as for the magnetization curve, discussed above, we find better agreement with the experimental data for the dipole-dipole interaction $\tilde{J}_0 = 3J_0$. The calculation revealed that the dipole-dipole interaction changes mostly the slope of the hysteresis curve near the values of magnetic field, where the magnetization vanishes (which are used to determine the coercive force iHc), rather than the coercive force itself, the latter is only weakly affected.

\begin{figure}[h]
	\includegraphics[width=\columnwidth]{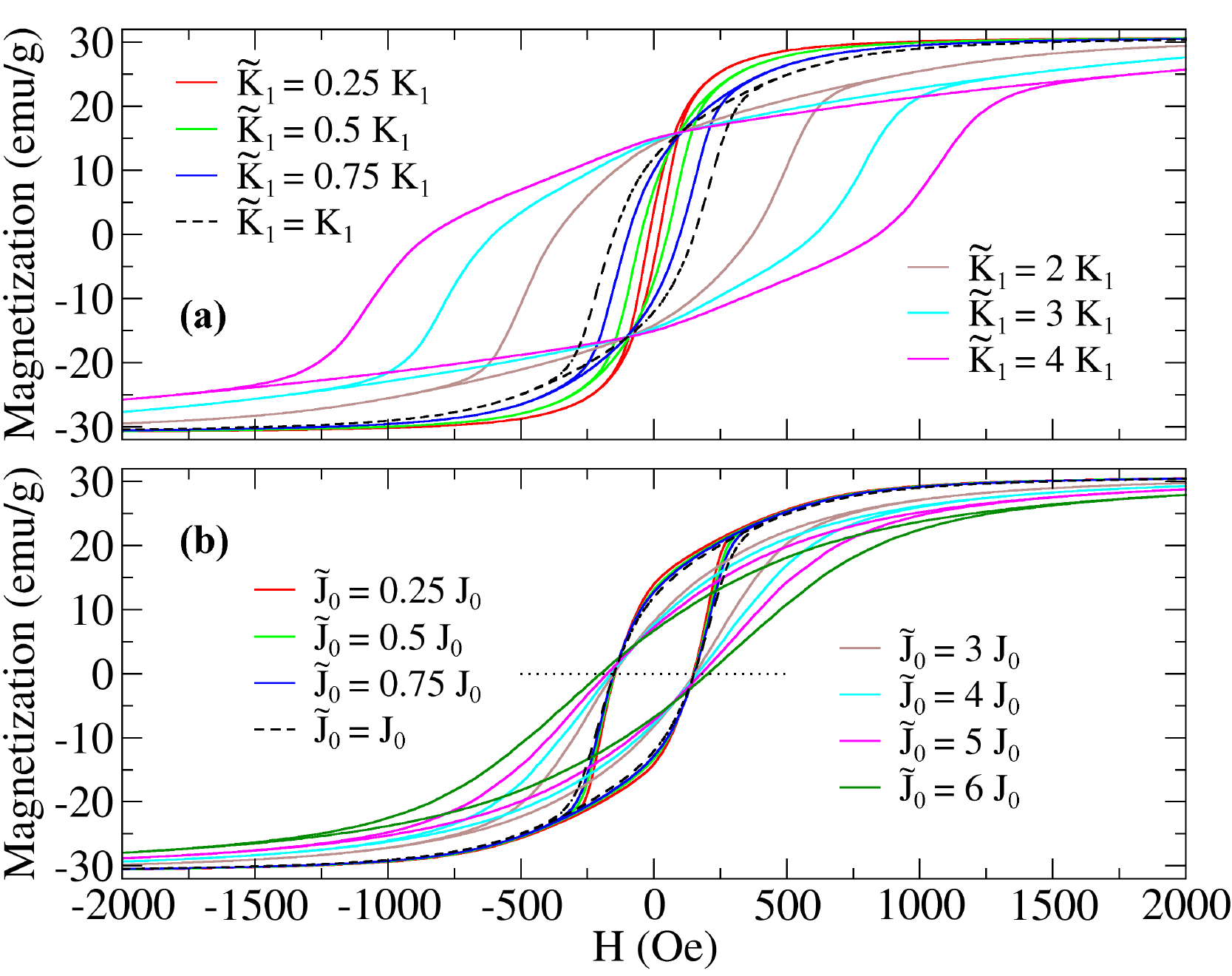}
	\caption{ (Color online) Dependence of the hysteresis loop profile on the (a) anisotropy constant $K_1$ and (b) dipole-dipole interaction constant $J_0$ model parameters. }\label{fig:hyst_var}
\end{figure}

To investigate the effect of changing parameters of the Hamiltonian on hysteresis loops, we plot in Fig.~\ref{fig:hyst_var} the results for various parameter sets. 
The anisotropy has strong impact on the hysteresis curves, yielding proportional change of the coercive force and fields which are required to reach saturated magnetization. At the same time, the dipole-dipole interaction yields only weak change of coercive force and changes mainly the slope of the hysteresis curve, as discussed above, and the magnetic field, which is necessary to reach saturation. Therefore, we find that the single-particle effects are more important for the shape of the hysteresis loops (in contrast to the ZFC magnetization curves studied above).

\begin{figure}[t]
	\includegraphics[width=\columnwidth]{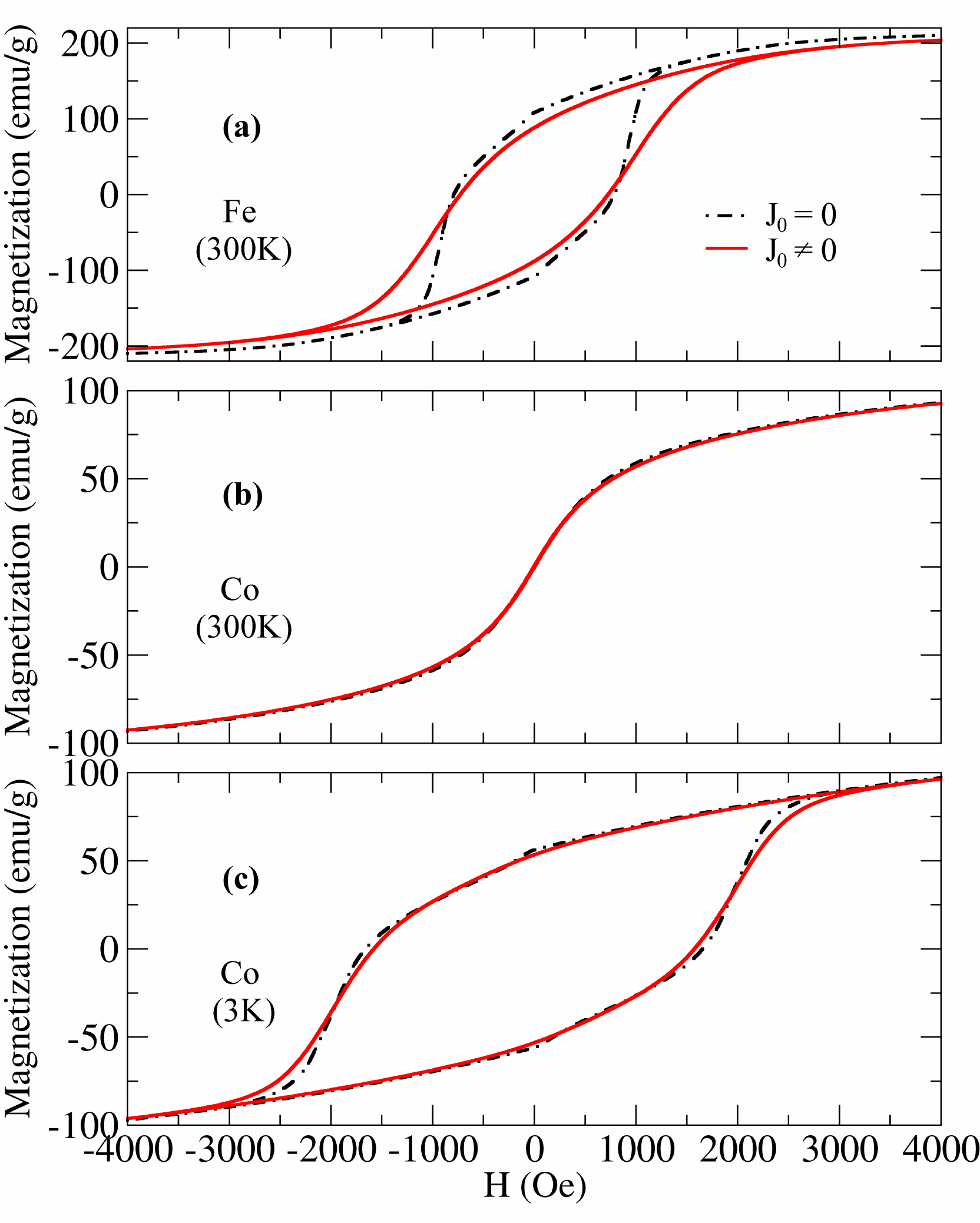}
	\caption {(Color online) Calculated hysteresis loops for (a) Fe- and (b), (c) Co nanoparticles adsorbed on the halloysite.
	}\label{fig:hyst_NiCoFe}
\end{figure}

On the basis of the performed calculations for Ni-nanoparticles adsorbed on the halloysite nanotubes, we carried out similar calculations for Fe and Co nanoparticles. The results of these simulations are presented in Fig.~\ref{fig:hyst_NiCoFe} and in Table~\ref{tab:results_sim}. One can see that Fe and Ni nanoparticles are magnetically hard in comparison with Co at 300 K, which appears to be in a superparamagnetic state. This result contradicts the experimental data from Ref.~\onlinecite{Co_halloy}, where adsorption of Co nanopatricles on the halloysite nanotubes surface was studied, and 
the hysteresis with rather large iHc of 1659 Oe was observed. At the same time, the authors of Ref.~\onlinecite{Co_pure} found out that Co nanoparticles are superparamagnetic at room temperature. Such a discrepancy in properties of Co nanoparticles can be attributed to difference in experimental techniques used for particles manufacturing in these papers. It is interesting to note that, according to our results (see Fig. \ref{fig:hyst_NiCoFe}~(c)), at low temperatures Co-modified halloysite exhibits a remnant magnetization and high coercive force (the latter is comparable to that obtained at room temperature in Ref. \onlinecite{Co_halloy}).

\begin{table}
\caption[Bset]{\label{tab:results_sim} Results on the hysteresis calculations in comparison with the experimental data.}
\renewcommand{\arraystretch}{1.7}
\begin{tabular}{cccc}
\hline
 Ion & $T$ (K) & iHc$^{\rm calc}$ (Oe) & iHc$^{\rm exp}$ (Oe)   \\
 \hline
 Ni & 300 &  151 &  110 
 [\onlinecite{Ni_halloy}]
 \\
 Fe & 300 &  727 & \--- \\
 Co & 300 &    5 & 1659 
 [\onlinecite{Co_halloy}]
 \\
 Co &   3 & 1577 & \---  \\
 \hline
\end{tabular}
\end{table}

\section{\label{sec:conclusion}Conclusion}
We have introduced a macroscopic spin Hamiltonian with dipolar interaction for simulating the magnetic properties of the cermet composite consisting of the Ni nanoparticles on the halloysite surface. The main distinctive features of our model which are the normally distributed sizes of the nanoparticles and the cylindrical lattice are aimed to the realistic reproduction of the conditions of the real experiment. 

Based on the results of the Monte Carlo simulations we conclude that the interaction of the dipole-dipole type between nanoparticles influences magnetic properties, yielding increase of the coercive force, and even stronger increase of the blocking temperature. Our calculations show a possibility of spin glass formation for the sufficiently strong dipole-dipole interaction.

A predictive modeling of hysteresis loops for the nanosystems with iron nanoparticles on the halloysite suggests a 
strong enhancement of the coercive force in comparison with the nickel ones. At the same time cobalt nanoparticles adsorbed on the nanotubes demonstrate superparamagnetic behavior at room temperature. 

We conclude that the proposed model allows to describe magnetic properties of the nanoparticles adsorbed on the cylindrical nanotubes. This study also raises the question of finding an `optimal' nanotubes to enhance the effects of the anisotropy and interparticle interactions, as well as motivates studying other experimental realizations of nanoparticles, adsorbed on different kinds of nanotubes.
For instance, it can be used for simulation of an ensemble of the nanoparticles loaded into inner spiral multi-walled structure of the halloysite. An experimental realization of such a loading was recently demonstrated in \cite{halloy_inner}. 

Other materials of technological importance such as carbon nanotubes decorated with metallic nanoparticles, for which numerous experimental data were collected over past 20 years \cite{carbon1,carbon2}, can be also modeled within the used approach. 
Within our model approach one can also control the anisotropy and transition to the superparamagnetic state at the nanoparticle size decreases, which is of primary interest in fields of ultrahigh-density recording and medicine \cite{Nature423}. 

\acknowledgments
We acknowledge fruitful discussions with Yuri Lvov,  Frederic Mila and Alexander Tsirlin.
This work was supported by the Russian Science Foundation, Grant 15-12-20021.

\appendix

\section{Sample geometry impact}
\label{sec:dipole_table}

This Appendix is devoted to discussing effects of sample surface curvature on the dipole-dipole interaction. Despite relatively large diameter of the tube, the surface curvature of cylindric lattice is significant due to the size of nanoparticles. Indeed, the cross-section of the tube with diameter 170 nm could comprise only up to twenty nanoparticles with size of 26.5 nm. 

\begin{figure}[h]
	\includegraphics[width=0.8\columnwidth]{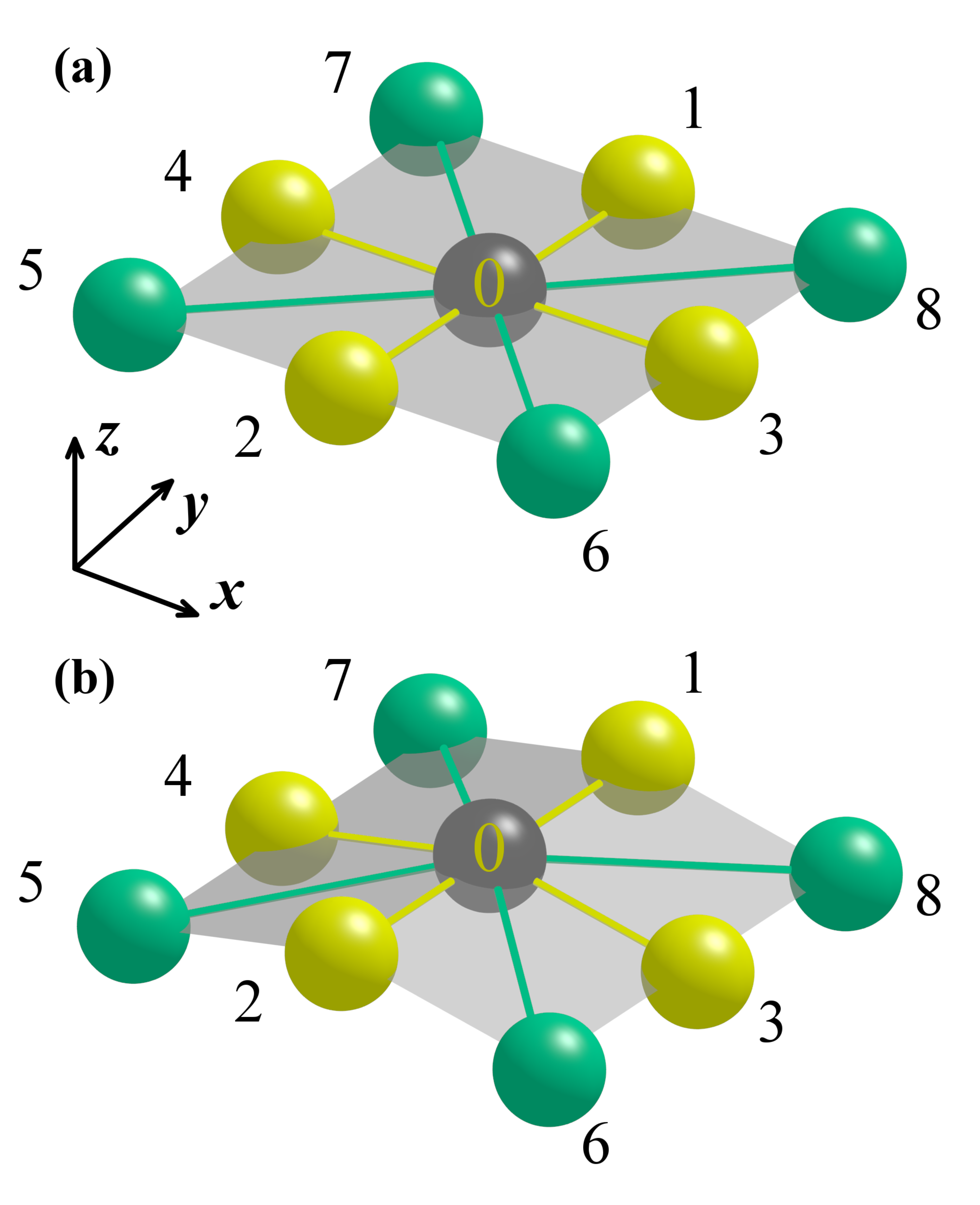}
	\caption {
    (Color online) Schematic representation of (a) flat and (b) cylindric lattice segments used in the calculation. The actual lattice constant and curvature radix are preserved. Particles size is reduced for clarity.}\label{fig:curvature_scheme}
\end{figure}

Fig.~\ref{fig:curvature_scheme} demonstrates segments of flat and cylindric square lattices involving a nanopatricle and its nearest- and next-nearest neighbors. It is clear, that rolled lattice possesses the extra spatial dimension in the transverse direction in comparison with flat topology. Taking into account anisotropic nature of dipole-dipole interaction~\eqref{eq:j}, this feature of cylindric lattice drastically changes the picture of pair interactions in the system. 

Table~\ref{tab:dipole_tensor} demonstrates the difference of lattice geometries in terms of dipole-dipole interaction tensors on the top of the tube. The positional indices ($ i $, $ j $) of $ J_{ij}^{\alpha\beta} $ correspond to site indices of Fig.~\ref{fig:curvature_scheme} whereas $ \alpha $, $ \beta = x,y,z$ components are represented as matrices in the table. For all considered neighbors, except $1$ and $2$ the curvature of the tube results in the 
additional non-diagonal \textit{xz}-components of the dipole-dipole interaction tensor. It is important to note, that absolute values of these non-diagonal $ J_{03,04}^{xz} $ components constitute about a one third of $ J_{03,04}^{xx} $ and more than a half of $ J_{03,04}^{yy} $ and $ J_{03,04}^{zz} $ diagonal components. Similar tendency can be observed for the second neighbors, although with smaller changes of the interaction tensor. 
The collective effect of such changes in pair interactions could be significant and considered in the main text of the paper. The changes can be even more drastic for smaller diameters of halloysite nanotubes 40-60 nm\cite{Lvov_review,halloysite_review}, which are not considered in the present study.


\begin{table}[h]
	\caption[Bset]{\label{tab:dipole_tensor} Comparison between square lattices of flat and cylindric shapes. The values of dipole-dipole tensors $J_{ij}^{\alpha\beta}$ are normalized by mean anisotropy $\braket{k} ={\pi}K_1\braket{d} ^3/{6}$. Changed components of tensors are marked with blue color. Upper and lower signs stand for first and second pair of indices.
    }
	\renewcommand{\arraystretch}{1.5}
	\begin{tabular}{c|c|c|}
		
		& Flat & Cylindric \\\hline
		
		$ J_{01,02}^{\alpha\beta} $
		&
		$ \begin{pmatrix}
		0.227 &  0  &  0 \\
		0  & -0.454   & 0 \\
		0  &  0  &  0.227
		\end{pmatrix} $
		&
		$ \begin{pmatrix}
		0.227  &  0  &  0 \\
		0 &  -0.454  &  0 \\
		0  &  0  &  0.227 
		\end{pmatrix} $
		\\\hline
		
		$ J_{03,04}^{\alpha\beta} $ 
		&   
		$ \begin{pmatrix}
		-0.454  &  0  &  0     \\
		0   & 0.227  &  0    \\
		0  &  0   & 0.227
		\end{pmatrix} $    
		&
		$ \begin{pmatrix}
		\hlmath{$-$0.431} &  0  & \hlmath{$\pm$0.123} \\
		0  &  0.227  &  0  \\
		\hlmath{$\pm$0.123} &  0  & \hlmath{0.204}
		\end{pmatrix} $ 
		\\\hline
        
		$ J_{05,08}^{\alpha\beta} $	
		&	 
		$ \begin{pmatrix}
		-0.04  & -0.12  &  0 		  \\
		-0.12  & -0.04  &  0 		  \\
		0 &   0  &  0.08
		\end{pmatrix} $ 		 
		&
		$ \begin{pmatrix}
		\hlmath{$-$0.036}  & \hlmath{$-$0.118} &  \hlmath{$\mp$0.022} \\
		\hlmath{$-$0.118} &  -0.04 & \hlmath{$\mp$0.022}  \\
		\hlmath{$\mp$0.022} &  \hlmath{$\mp$0.022}  &  \hlmath{0.076}
		\end{pmatrix} $ 
		\\\hline
		
		$ J_{06,07}^{\alpha\beta} $ 
		&   
		$ \begin{pmatrix}
		-0.04  &  0.12   & 0    \\
		0.12  & -0.04  &  0    \\
		0  &  0  &  0.08
		\end{pmatrix} $ 
		&
		$ \begin{pmatrix}
		\hlmath{$-$0.036}  &  \hlmath{0.118}  &  \hlmath{$\pm$0.022} \\
		\hlmath{0.118} &  -0.04  & \hlmath{$\mp$0.022} \\
		\hlmath{$\pm$0.022} &  \hlmath{$\mp$0.022} &   \hlmath{0.076}
		\end{pmatrix} $
		\\\hline
        
	\end{tabular}
\end{table}

\section{Solver Adjustment}
\label{sec:appendix}

In this Appendix we provide the details of Monte-Carlo algorithm.

For initial system state
preparation a heat bath\cite{Landau} algorithm with 10000 heat bath sweeps was used. Then additionally $10000$ SAR Monte Carlo sweeps without accumulation of the desired averages were performed. The single SAR sweep consists of $N$ elementary spin rotations, where $N$ is the total number of the lattice sites. After each rotation the new state of the system is accepted with probability $exp(-\beta\Delta{}E)$, where $\beta = {1}/{T}$ is the inverse temperature ($T$ is measured in energy units) and $\Delta{}E$ is the energy change due to the spin rotation.

This thermalized sample is then treated in the way, depending on the type of the simulated magnetization curve. For the {\it hysteresis loops} we choose the fixed temperature and thermalize the system at the initial field $H$. Then gradually increasing magnetic field with step $\Delta{}H$ we perform $N_{\rm SAR}$ Monte Carlo sweeps for each field step. The reverse magnetization curve is calculated in the exact same way, starting from the final spin configuration of the forward magnetization curve.

In the case of the {\it ZFC magnetization curves}, the initial state is thermalized at $H = 0$ and $T_{\rm max}=3000$ K. Then the system has been slowly cooled down. During this process we were gradually decreasing the temperature with step $\Delta{}T$, performing $N_{\rm SAR}$ Monte Carlo sweeps without accumulation of averages at each temperature step. After that, we set a relatively small magnetic field $H$ and heated the system in a similar way calculating the average value of the magnetization. The solid angle restriction $\Delta \Omega$ for thermalization and Metropolis scheme has been chosen to get an appropriate acceptance rate during the Monte Carlo simulation \cite{MC_timestep,pereiera2004}. 

The observable quantities are calculated during the simulation in the standard way. For instance, the magnetization is defined as
\begin{equation}
\label{eq:averaging}
\braket{M} = \frac{1}{N}\sum_{n=1}^N{M_n}
\end{equation}
where $M_n$ represents the projection of the total magnetization ($M_n = \sum_{i}\mu_i\mathbf{S}_i^{(n)}\cdot\mathbf{h}$) of the system on $n$-th Monte Carlo spin-flip attempt on the magnetic field direction $\mathbf{h}$ and $N$ denotes the total number of the flip attempts within a single Monte Carlo sweep.

\begin{figure}
	\includegraphics[width=0.9\columnwidth]{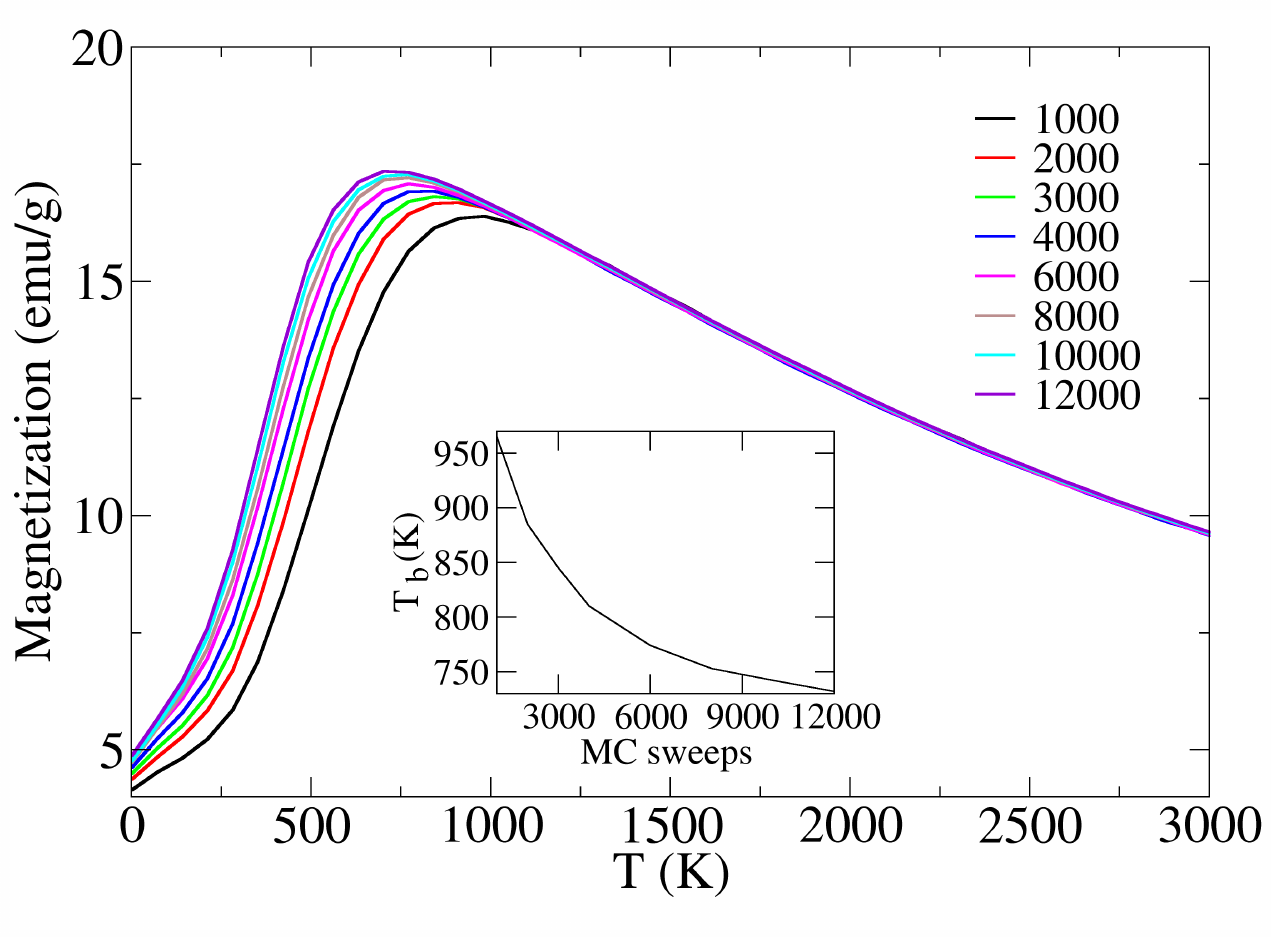}
	\caption {(Color online) ZFC magnetization curves for different number of SAR Monte Carlo steps at $J_0=0$ as indicated in the legend. The maximum of ZFC magnetization curves corresponds to blocking temperature $T_b$. The inset shows the dependence of the blocking temperature on the number of Monte-Carlo steps. }\label{fig:zfc_sar}
\end{figure}

It is important to note that treatment of the partially occupied lattice actually models a disordered system. Therefore one has to perform a configurational averaging:
\begin{equation}
\label{eq:conf_averaging}
\braket{\braket{M}} = \frac{1}{N_C}\sum_{i=1}^{N_C}\braket{M}_i
\end{equation}
where $\braket{M}_i$ corresponds to the average value of the magnetization calculated as Eq.~\eqref{eq:averaging} for a distinct configuration and $N_C$ denotes the total number of the generated initial configurations in the way described above. Depending on the type of the experiment we would like to describe, the average values Eqs.~\eqref{eq:averaging} and~\eqref{eq:conf_averaging} that present ZFC magnetization curves 
$\braket{\braket{M}}(T)$ or hysteresis magnetization loops 
$\braket{\braket{M}}(H)$.

The configurational averaging is of crucial importance when the system is characterized by relatively small number of the occupied sites, which is the case here. It is clear that fluctuations mostly appear in the high temperature region. Thus, one has to set a relatively large number $N_C$ of the configurational averages for the calculations of the ZFC magnetization curves. At the same time hysteresis loops are usually calculated at room temperature and below. Therefore it is possible to reduce $N_C$ in the case of the hysteresis modeling. In our simulations we set $N_C = 200$ for ZFC magnetization curves and $N_C = 120$ for hysteresis loops calculation.

\begin{figure}[t]
\includegraphics[width=\columnwidth]{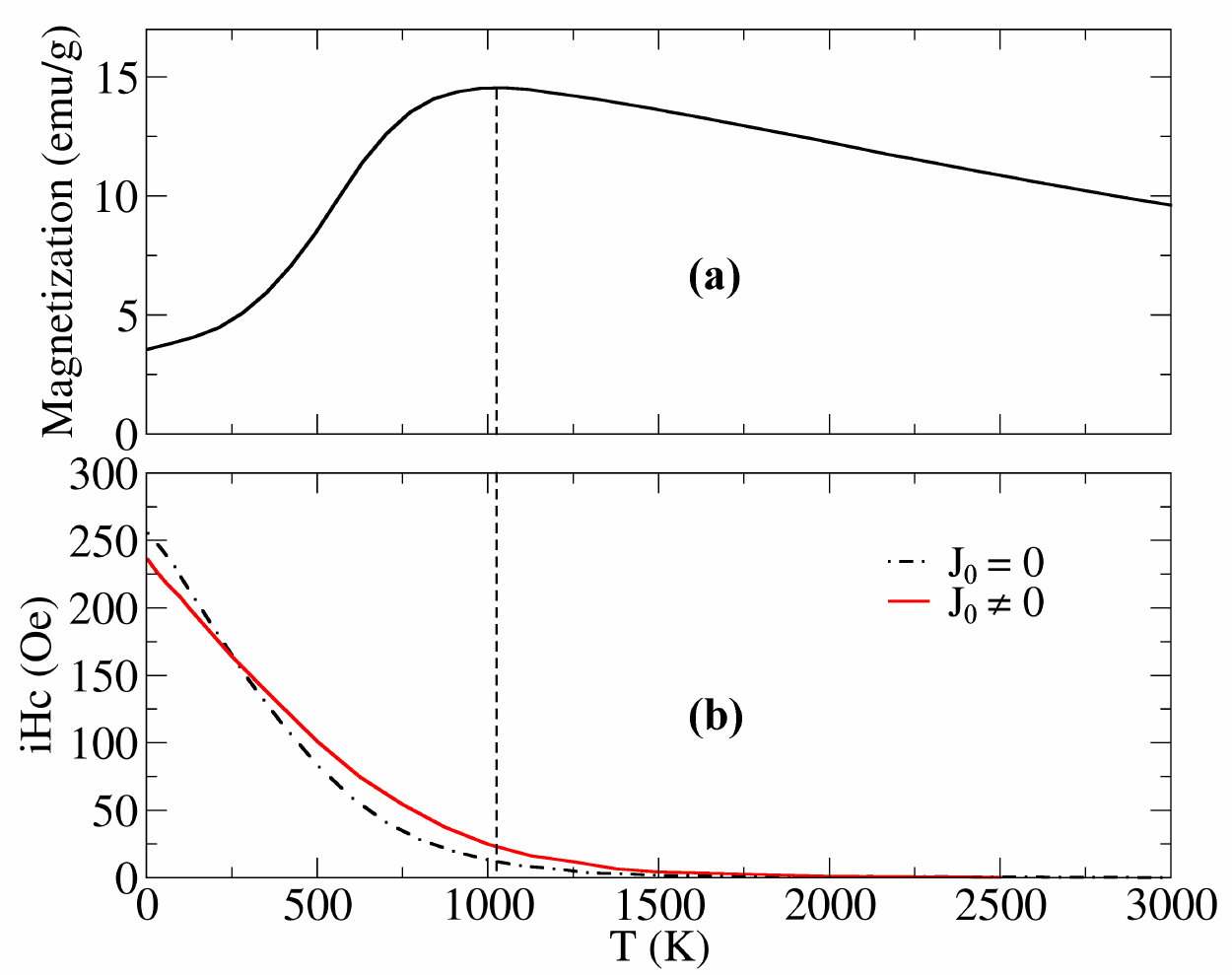}
\caption {(Color online) (a) ZFC magnetization curve, calculated with $\Delta{}T = 70$ K, $ N_{SAR} = 6000$ (b) Coercive force as function of temperature obtained with $\Delta{}H = 10 $ Oe, $N_{SAR} = 2000$ for Ni-modified halloysite. The concentration of the particles is equal to 0.6 for both curves. }\label{fig:coer_Ni}
\end{figure}

The choice of the SAR Monte Carlo simulation parameters  affects the resulting curves, for example too large value of $N_{\rm SAR}$ will lead to the situation when the system falls into the ground state at each temperature or field step, which results in narrowing of the hysteresis and decreasing of the blocking temperature $T_b$. This can be attributed to the problem of the correspondence of the Monte Carlo simulation timescale to that of the real experiment, which makes quantitative description of experimental data quite hard. The preparation of the initial state and averaging techniques are also important in this context. 

Figure~\ref{fig:zfc_sar} demonstrates the typical behavior of the zero-field cooled magnetization curves on the number of the Monte Carlo sweeps per temperature step for the non-interacting nanoparticles system ($J_0 = 0$). One can observe the decrease of the  blocking temperature with increase of the number of $N_{\rm SAR}$ sweeps. This decrease is consistent with the previously reported one~\cite{pereiera2004}, and also with the Neel formula for the blocking temperature of non-interacting nanoparticles of equal size, $T_b=K_1 V/\ln(\tau/\tau_0)$, where $\tau$ is the waiting (or measurment) time, and $\tau_0$ is the characteristic time scale, which is determined by concrete experimental realization, or, in our case, the details of the Monte Carlo calculation.

In order to reproduce the experimental hysteresis loop, we use the following procedure. Having fixed $\Delta\Omega=0.7$, we are to define $\Delta H$, $\Delta T$ and $N_{SAR}$. The parameters $\Delta H$ and $N_{SAR}$ can be estimated from the fitting of the experimental magnetization curves. We have found that the values $\Delta H = {10}$~Oe and $N_{SAR} = {2000}$ allow to fit the low- and high-field parts of the magnetization process in the case of the sample, which was heated up to 573 K and then cooled to the room temperature in Ref. \onlinecite{Ni_halloy}, as depicted in Fig.~\ref{fig:m573K}.

The value $\Delta T={70}$ K of the temperature step for these curves was chosen in such a way, that the blocking temperature, obtained from ZFC magnetization curves coincides with that, obtained from the condition of almost vanishing coercive force ${\mathrm {iHc}}
(T_b)\approx 0$, see Fig.~\ref{fig:coer_Ni}. From Fig.~\ref{fig:coer_Ni}~(b) one can also see that the dipole-dipole interaction generally leads to increase of the coercive force, except the very low temperatures, where the opposite tendency is observed. The latter can be however the artifact of the used Monte Carlo method, since this low-temperature regime may require more complicated treatment of the spin angle restriction etc.

\end{document}